\documentstyle[aps,multicol,prl]{revtex}
\draft
\begin{document}
\title{Local fractional Fokker-Planck equation}
\author{Kiran M. Kolwankar\cite{emk} and Anil D. Gangal\cite{emg}}
\address{Department of Physics, University of Pune, Pune 411 007 India}
\maketitle
\centerline{(April 2 1997)(cond-mat/9801138)}
\begin{abstract}
New kind of differential equations, called local fractional differential 
equations,
has been proposed for the first time.
They involve local fractional derivatives
introduced recently. Such equations appear to be suitable
 to deal with phenomena taking place in fractal space and time.
A local fractional analog of 
Fokker-Planck equation has been derived starting from
the Chapman-Kolmogorov condition.
Such an equation is solved, with a specific choice of the transition 
probability, and shown to give rise to subdiffusive behavior.
\end{abstract}
\begin{multicols}{2}
Derivatives and integrals of fractional order have found many applications
in recent studies of scaling phenomena~\cite{Non,GN,Sch,MV}. 
The main aim of the most of these papers
is to formulate  fractional integro-differential equations to
describe some scaling process. Modifications of equations governing physical processes such as
 diffusion equation, wave equation and Fokker-Planck equation have been
suggested~\cite{GR,RG,Wys,SW,Jum,Fog} which incorporate fractional derivatives 
 with respect to time.
Recently Zaslavasky~\cite{Zas} argued that the chaotic Hamiltonian  dynamics
of particles can be described by using fractional generalization of the
Fokker-Planck-Kolmogorov (FPK) equation.
However fractional derivatives are nonlocal and hence such equations are not
suitable to study local scaling behavior.   In the present work  we
rigorously derive fractional analogs of equations like 
the FPK equation involving one space variable. 
Our approach differs from the above mentioned
ones since we use local fractional Taylor expansion, which was established
only recently \cite{KG1}. As is argued below, such equations can provide
appropriate  schemes for describing evolutions (e.g. sub or 
superdiffusive) normally not obtained from the usual FPK
equation.

 It was realized  recently \cite{KG1}
 that there is a direct quantitative connection
between fractional differentiability properties of continuous 
but nowhere differentiable  functions and the dimensions of their graphs.
 In order to show this, a 
 new notion of local fractional derivative (LFD) was introduced. 
The LFD of order $q$ of a function $f(y)$ was defined by
\begin{eqnarray}
I\!\!D^qf(y) =  {\lim_{x\rightarrow y}}
{{d^q[f(x)-f(y)]}
\over{[d(x-y)]^q}} \;\;\;0 < q \leq 1\label{deflocg}
\end{eqnarray}
where the derivative  on the RHS is the
 Riemann-Liouville fractional derivative \cite{OS,RM}, viz., for $0<q<1$
\begin{eqnarray}
{{d^qf(x)}\over{[d(x-a)]^q}}={1\over\Gamma(1-q)}{d\over{dx}}{\int_a^x{{f(y)}
\over{(x-y)^{q}}}}dy.\label{def2}
\end{eqnarray}
As is obvious from equation~(\ref{def2}), the operator 
${d^q} / {[d(x-a)]^q}$ is nonlocal and further the
${d^qf(x)} / {[d(x-a)]^q} \neq 0$ 
for $f(x)=\mbox{Const}$. 
The motivation for the definition of $I\!\!D^q$ was to correct
for both of these features.
 It was shown in \cite{KG1},
in particular, that the LFD of Weierstrass nowhere differentiable function 
exists upto (critical) order $1-\gamma$, 
where $1+\gamma$ ($0<\gamma<1$) is the box dimension 
of the graph
of the function. Further the use of LFD to study pointwise behavior of
multifractal functions was also demonstrated.
The definition was generalized~\cite{KG2} for a function for which
the first $N$ derivatives exist by replacing
$[f(x)-f(y)]$ on RHS of equation (\ref{deflocg}) by
\begin{eqnarray}
\widetilde{F}_N(x,y) = f(x)-\sum_{n=0}^N{f^{(n)}(y)\over\Gamma(n+1)}(x-y)^n
\end{eqnarray}
with $q$ in the interval $(N, N+1]$.
Sometimes it is essential to distinguish between limits 
taken from above and below. In that case we define
\begin{eqnarray}
I\!\!D_{\pm}^qf(y) =  {\lim_{x\rightarrow y^{\pm}}}
{{d^q\widetilde{F}_N(x,y)}
\over{[d\pm(x-y)]^q}}. \label{deflocg+}
\end{eqnarray}
We will assume $I\!\!D^q = I\!\!D_{+}^q$ unless mentioned otherwise.
We note that when $q=n$, an integer, $I\!\!D^qf(y)$ is simply $d^nf(y)/dy^n$.

The importance of the above definition also lies in the fact that
the LFDs appear naturally in the fractional Taylor expansion 
as the coefficient of the power with fractional exponent.
Thus, for $\Delta=x-y$
\begin{eqnarray}
f(x)\! = \!\sum_{n=0}^{N}{f^{(n)}(y)\over{\Gamma(n+1)}}\Delta^n \!
 + {I\!\!D_{\pm}^qf(y)\over \Gamma(q+1)} (\pm\Delta)^q + R_q(y,\Delta) 
\label{taylor2}
\end{eqnarray}
where $R_q(y,\Delta)$ is the remainder~\cite{KG1}.

The basic idea of the present paper is to utilize such fractional
Taylor expansions in
the Chapman-Kolmogorov condition and obtain analogs of FPK equation.
We begin by recalling the usual procedure and difficulties of obtaining
FPK equation.
Let $W(x,t)$ denote the probability density for a random variable
$X$ taking value $x$ at time $t$, then
\begin{eqnarray}
W(x,t+\tau)=\int P(x,t+\tau | x',t) W(x',t) dx' \label{CK}
\end{eqnarray}
where  $P(x_1,t_1 | x_2,t_2)$
denotes the transition probability from $x_1$ at time 
$t_1$ to $x_2$ at time $t_2$
and $\tau \geq 0$. The usual FPK equation is obtained~\cite{Ris} from 
equation~(\ref{CK}) by expanding the integrand in a Taylor series.

There are number of limitations of this approach arising naturally
from the assumptions going into its derivation. For instance, 
as noted in \cite{LM}, probability
distributions whose second moment does not exist are not described by FPK 
equation even though such distributions may satisfy original 
Chapman-Kolmogorov equation.
Also, as emphasized in \cite{Fel2}, the differentiability assumption may also 
break down in various situations. For instance, the transitional probability
density may not be differentiable at $x=x'$ in which case the derivation
of FP equation itself will break down. Another situation is when
we have a fractal function as the initial probability density. In such a case
even the usual Fokker-Planck operator can not be operated on the initial 
density. 

It is thus of interest to broaden the class of differential  
equations one can derive starting from the Chapman-Kolmogorov equation, 
and study various 
processes described by them. In this paper we pursue the possibility
of removing the assumption of differentiability of probability densities.
We follow the usual procedure to derive the  
Fokker-Planck equation from equation~(\ref{CK}) except that we now expand
the integrand using
fractional Taylor expansion~(\ref{taylor2}) instead of 
ordinary Taylor expansion. Thus, if $\Delta = x-x'$,
\end{multicols}
\vrule width3.375in height.2pt depth.2pt \vrule  depth0em height1em \hfill
\begin{eqnarray}
W(x,t+\tau)&=&W(x,t) +\sum_{n=1}^N {1\over\Gamma(n+1)}({\partial\over{\partial(-x)}})^n
\int dx' \Delta^n P(x+\Delta,t+\tau | x,t) W(x,t)\nonumber \\
&&+ {1\over\Gamma(\beta+1)}
I\!\!D_{x-}^{\beta}[\int_x^{\infty} dy {(y-x)}^{\beta} 
P(y,t+\tau | x,t) W(x,t)]\nonumber\\
&&\;\;\;+ {1\over\Gamma(\beta+1)}
I\!\!D_{x+}^{\beta}[\int_{-\infty}^x dy {(x-y)}^{\beta} 
P(y,t+\tau | x,t) W(x,t)]+ Remainder
\end{eqnarray}
\hfill\vrule depth1em height0pt \vrule width3.375in height.2pt depth.2pt
\begin{multicols}{2}
\noindent
where $I\!\!D_x$ is a partial LFD w.r.t $x$.
Now if $0< \alpha \leq 1$
\begin{eqnarray}
W(x,t+\tau)-W(x,t) = {\tau^{\alpha}I\!\!D_t^{\alpha}W(x,t)\over\Gamma(\alpha+1)}
+ Remainder. \nonumber
\end{eqnarray}
where $I\!\!D_t$ is partial LFD w.r.t. $t$.
In general $\alpha$ and $\beta$ may depend on $x$ and $t$.  But we 
assume that $\alpha$ and $\beta$ are constants.
Therefore we get
\begin{eqnarray}
{\tau^{\alpha}I\!\!D_t^{\alpha}W(x,t)\over\Gamma(\alpha+1)} &=&
\sum_{n=1}^N \big({\partial\over{\partial(-x)}}\big)^n
\big[{M_n(x,t,\tau)\over\Gamma(n+1)}W(x,t)\big] \nonumber \\ 
&&+ 
I\!\!D_{x-}^{\beta}\big[{M_{\beta}^+(x,t,\tau)\over\Gamma(\beta+1)}W(x,t)\big]
\nonumber \\ 
&& + 
I\!\!D_{x+}^{\beta}\big[{M_{\beta}^-(x,t,\tau)\over\Gamma(\beta+1)} W(x,t)\big]
\end{eqnarray}
where
\begin{eqnarray}
M_a^+(x,t,\tau)=\int_x^{\infty} dy (y-x)^a P(y,t+\tau|x,t) \;a>0,
\end{eqnarray}
\begin{eqnarray}
M_a^-(x,t,\tau)=\int_{-\infty}^x dy (x-y)^a P(y,t+\tau|x,t) \;a>0
\end{eqnarray}
and
\begin{eqnarray}
M_a(x,t,\tau)=M_a^+(x,t,\tau)+M_a^-(x,t,\tau) \label{eq:moments}
\end{eqnarray}
are transitional moments.
The limit $\tau \rightarrow 0$ gives us a equation
\begin{eqnarray}
I\!\!D_t^{\alpha}W(x,t) \equiv {\cal{L}}(x,t)W(x,t) \label{trunKM}
\end{eqnarray}
where the operator $\cal{L}$ is given by
\begin{eqnarray}
{\cal{L}}(x,t)=&& 
\sum_{n=1}^N \big({\partial\over{\partial(-x)}}\big)^n
A_{\alpha}^n(x,t)  +
I\!\!D_{x-}^{\beta}A_{\alpha -}^{\beta}(x,t) \nonumber \\ &&+
I\!\!D_{x+}^{\beta}A_{\alpha +}^{\beta}(x,t)  
\end{eqnarray}
where
\begin{eqnarray}
A_{\alpha \mp}^{\beta}(x,t)= \lim_{\tau \rightarrow 0}
{M_{\beta}^\pm(x,t,\tau)\Gamma(\alpha+1)\over{\tau^{\alpha}\Gamma(\beta+1)}}
\label{mbeta}
\end{eqnarray}
and
\begin{eqnarray}
A_{\alpha }^{\beta}(x,t)=A_{\alpha +}^{\beta}(x,t)+A_{\alpha -}^{\beta}(x,t).
\end{eqnarray}
Here corresponding $A_{\alpha}$'s are assumed to exist.
We would like to point out that the equation (\ref{trunKM}) is analogous to 
truncated Kramers-Moyal expansion. Two rather important special cases are,
 $0<\beta<1$ and $1< \beta < 2$. In the former case we get the operator
\begin{eqnarray}
{\cal{L}}(x,t) &=&
I\!\!D_{x-}^{\beta}A_{\alpha -}^{\beta}(x,t)+ 
I\!\!D_{x+}^{\beta}A_{\alpha +}^{\beta}(x,t), \nonumber
\end{eqnarray}
and  in the latter case we get
\begin{eqnarray}
{\cal{L}}(x,t)&=&-{\partial\over{\partial x}}
A_{\alpha}^1(x,t)\!+\!I\!\!D_{x-}^{\beta}A_{\alpha -}^{\beta}(x,t)\!+\! 
I\!\!D_{x+}^{\beta}A_{\alpha +}^{\beta}(x,t) \nonumber
\end{eqnarray}
This operator can be identified as   generalizations of the Fokker-Planck 
operator in one space variable. 
It is clear that when $\alpha=1$ and $\beta=2$ we get back the
usual Fokker-Planck operator.

It may be pointed out that the local fractional differential equations (LFDE)
that we are proposing here are new kind of differential equations. To our
knowledge this is the first direct occurrence of such equations.  
We note that they are different from the conventional fractional differential
equations which have been studied to some extent in the literature \cite{OS,RM}
and which have found several applications ranging from solutions of Bessel equation, diffusion on curved surfaces to wave equation etc. In fact the
equations appearing in \cite{Non,GN,GR,RG,Wys,SW,Jum,Zas}
 are all conventional fractional differential
equations. On the other hand the present LFDE involve operators $I\!\!D^q$,
which found successful applications \cite{KG1} in studying differentiability
properties of nowhere differentiable functions and relating them to dimensions.
They are appropriate to address scaling phenomena. It is for this reason that 
one would expect the equations governing the fractal processes to be LFDE. At
this stage it is worth reflecting for a moment 
on the behavior of meaningful solutions
of simple LFDE. We begin by considering the equation
\begin{eqnarray}
I\!\!D^q_xf(x) = g(x). \label{eq:slfde}
\end{eqnarray}
The questions of the general conditions guaranteeing solutions of such an 
equation is an involved one.
We note that the equation $I\!\!D^q_xf(x)= \mbox{Const}$, does not have
a finite solution when $0<q<1$. Interestingly, the solutions to (\ref{eq:slfde})
can exist, when $g(x)$ has a fractal support. For instance, when 
$g(x)=\chi_C(x)$, the membership function of a cantor set $C$
(i.e. $g(x)=1$ if $x$ is in $C$ and $g(x)=0$ otherwise), the solution with
initial condition $f(0)=0$ exists if $q=\alpha \equiv \mbox{dim}_HC$.
Explicitly, generalizing the Riemann integration procedure, 
\begin{eqnarray}
f(x)\equiv { P_C(x)\over\Gamma(\alpha+1)}
=\lim_{N\rightarrow \infty} \sum_{i=0}^{N-1} {(x_{i+1}-x_i)^\alpha
\over\Gamma(\alpha+1)} F_C^i \label{eq:PCt}
\end{eqnarray}
where $x_i$ are  subdivision points of the interval $[x_0=0, x_N=x]$ and
$F_C^i$ is a flag function which takes value 1 if the interval $[x_i,x_{i+1}]$
contains a point of the set $C$ and 0 otherwise. 
Note that $P_C(x)$ is a Lebesgue-Cantor 
(staircase) function and satisfies the bounds 
$ax^\alpha \leq P_C(x) \leq bx^\alpha$ where $a$ and  $b$
are suitable positive constants. 
In general, the algorithm of the equation
(\ref{eq:PCt}) will work only for the sets $C$ for which 
$\mbox{dim}_BC=\mbox{dim}_HC$ (in fact in this case only $N^\alpha$
terms in the summation are nonzero).
More details about solutions of such equations 
and algorithms will
be discussed elsewhere \cite{KG3}.

Returning back to equation (\ref{mbeta}) it is clear that  the small time
behavior of different transitional moments decide the order of the 
derivative with respect to time ( in order to demonstrate this point we
consider the example of a L\'evy process below).
On the other hand, small distance behavior of transitional probability or
the differentiability property of the initial probability density would
dictate the order of space derivative.
Depending on the actual values of 
$\alpha$ and $\beta$ as well as their interrelation the above local FFPK
equation will describe different processes. 

Equations which give rise to evolution-semigroup are of interest in physics.
The equation~(\ref{trunKM}) corresponds to a semigroup if $\alpha=1$. 
One can then write down a formal solution of the above equation in this case as
follows. In the time independent case we have
\begin{eqnarray}
W(x,t) = e^{{\cal{L}}(x) t}W(x,0)
\end{eqnarray} 
and  when the operator depends on time we have
\begin{eqnarray}
W(x,t) = \stackrel{\leftarrow}{T}e^{\int_0^t{\cal{L}}(x,t') dt'}W(x,0)
\end{eqnarray} 
where ${\cal{L}}$ is an operator in equation (\ref{trunKM}) and
 $\stackrel{\leftarrow}{T}$ is the time ordering operator.

For the symmetric stable L\'evy process of index $\mu$, the moments scale as
$M_\gamma(\lambda t) = \lambda^{\gamma /\mu} M_{\gamma}(t)$
and we get
\begin{eqnarray}
I\!\!D_t^{\gamma/ \mu}W(x,t)= &&
I\!\!D_{x-}^{\gamma}[A_{\gamma/ \mu -}^{\gamma}(x,t) W(x,t)]\nonumber\\&&+ 
I\!\!D_{x+}^{\gamma}[A_{\gamma/ \mu +}^{\gamma}(x,t) W(x,t)]. \label{FPL}
\end{eqnarray}
Since the process is symmetric the first derivative does not appear.
The order of the time derivative depends on that of space derivative but
it is always less than one. Now there is only one free parameter $\gamma$
which is restricted to the range $0 <\gamma <\mu$. 
In this case the value of $\gamma$ will be decided by the differentiability
class of the initial distribution function. 
(The details and intricacies will be addressed in~\cite{KG3}.)
When $\mu = 2$ and 
$\gamma = 2$ we get back the usual Fokker-Planck equation describing
a Gaussian process. Equation (\ref{FPL}) forms one example where usual 
derivation of FPK equation 
breaks down and we get nontrivial values for the orders of the
derivatives.

As our next example we consider the transition probability 
\begin{eqnarray}
P(x,t+\tau|x',t) &=& {1\over\sqrt{\pi\Delta{P}_C(t,\tau)}} 
 e^{-(x-x')^2\over{\Delta{P}_C(t,\tau)}} \\
&=& \delta(x-x') \;\;\;\mbox{if} \;\;\;\Delta{P}_C(t,\tau) = 0,
\end{eqnarray}
where $\Delta{P}_C(t,\tau)=P_C(t+\tau)-P_C(t))$.
This transition probability describes a nonstationary process which corresponds 
to transitions occouring only at times which
lie on a fractal set. Such a transition probability can be used to model
phenomenon where transition is very rare, for instance, diffusion in the 
presence of traps.
The second moment is given, from equation~(\ref{eq:moments}), by
\begin{eqnarray}
M_2(t,\tau) &=& {\Delta{P}_C(t,\tau)\over{2}} 
\simeq {1\over 2}{ I\!\!D^\alpha P_C(t) \over \Gamma(\alpha+1)} \tau^\alpha 
\nonumber\\
&=& {\tau^{\alpha}\over{2}} \chi_C(t)
\end{eqnarray}
This gives us the following local fractional Fokker-Planck equation 
(in this case an analog of a diffusion equation).
\begin{eqnarray}
I\!\!D_t^{\alpha}W(x,t) 
&=& {\Gamma(\alpha + 1)\over 4} \chi_C(t) {\partial^2\over{\partial x^2}}W(x,t)
\label{eq:ex} 
\end{eqnarray}
We note that even though the variable $t$ is taking all real positive values
the actual evolution takes place only for values of $t$ in the fractal set $C$.
The solution of equation~(\ref{eq:ex}) can easily be obtained as
\begin{eqnarray}
W(x,t) 
&=& P_{t-t_0}W(x,t_0)
\end{eqnarray}
where
\begin{eqnarray}
P_{t-t_0}
&=&  \lim_{N\rightarrow \infty} \prod_{i=0}^{N-1} \big[ 1 + {1\over 4} (t_{i+1}-t_{i})^\alpha F_C^i {\partial^2\over{\partial
x^2}}\big].
\end{eqnarray}
The above product converges because except for number of terms of order 
$N^{\alpha}$ all other terms
take value 1.
It is clear that for $t_0 < t' < t$
\begin{eqnarray}
W(x,t) = P_{t-t'}P_{t'-t_0}W(x,t_0)
\end{eqnarray}
and $P_t$ gives rise to a semigroup evolution.
Using equation~(\ref{eq:PCt}) it can be 
easily seen that
\begin{eqnarray}
W(x,t) = e^{{P_C(t)\over 4}{\partial^2\over{\partial
x^2}}}W(x,t_0=0).
\end{eqnarray}
Now  choosing $W(x,0) = \delta(x)$ and using the
 Fourier representation
of delta function then we get the solution
\begin{eqnarray}
W(x,t) &=& {1\over\sqrt{\pi P_C(t) }}e^{ - x^2\over{P_C(t)}}
\end{eqnarray}
Its consistency can easily be checked by substituting this in Chapman-Kolmogorov
equation.
This solution satisfies the bounds
\begin{eqnarray}
{1\over\sqrt{\pi bt^\alpha }}e^{ - x^2\over{bt^\alpha}} \leq 
W(x,t)  \leq
{1\over\sqrt{\pi at^\alpha }}e^{ - x^2\over{at^\alpha}}
\end{eqnarray}
for some $0<a<b$.
This is a model solution of a subdiffusive behavior. It is clear that
when $\alpha=1$ we get back the typical solution of the ordinary
diffusion equation which is $(\pi t)^{-1/2}\exp(-x^2/t)$.

To conclude, we have derived the generalization of FP equation
which involves the local fractional derivatives. 
Our equations are fundamentally different
from any of the equations proposed previously
since they involve LFDs. They are  local and
more natural generalization of ordinary differential equations.
LFDEs deserve a separate study in their own right~\cite{KG3}.
We would like to point out that LFDEs naturally give rise to dynamical
systems of new kind (neither discrete nor continuous) in which time evolution
takes place for values of time belonging to a Cantor-like set.
We further remark that since the present analogue of FPK equation is derived 
from first principles, we feel that our equation will have general
applicability in the field of physics. We expect them to be of value
in the studies of anomalous diffusion, chaotic Hamiltonian systems, 
disordered phenomenon, etc.
In our derivation we assumed that the orders $\alpha$ and $\beta$ of derivatives involved are constants. This would require a modification
the description of multiscaling
multifractal processes. We further note that directional LFDs are defined
in \cite{KG4}. Using them it may be possible to obtain local FPK equation
involving several variables.

We would like to thank Prof. J. Kupsch, Dr. H. Bhate and Dr. U. 
Naik-Nimbalkar for
helpful discussions. One of the author (KMK) would like to thank 
CSIR (India) and the other author (ADG) would 
like to thank DBT (India) for financial assistance.

\end{multicols}

\begin{thebibliography}{abc}
\bibitem[*]{emk} Electronic address: kirkol@physics.unipune.ernet.in
\bibitem[\dag]{emg} Electronics address: adg@physics.unipune.ernet.in
\bibitem{Non} T. F. Nonnenmacher,  {\it J. Phys. A: Math. Gen.}
 {\bf 23}, L697 (1990).
\bibitem{GN} W. G. Gl\"ockle and T. F. Nonnenmacher, {\it J. Stat. Phys.} 
{\bf 71}, 741 (1993).
\bibitem{Sch} M. F. Schlesinger, {\it J. Stat. Phys.} {\bf 36}, 639 (1984).
\bibitem{MV} B. B. Mandelbrot and J. W. Van Ness, {\it SIAM Rev.} {\bf 10}, 422 
(1968).
\bibitem{GR} M. Giona  and H. E. Roman,  {\it J. Phys. A: Math Gen.} {\bf 25}, 
2093 (1992) 
\bibitem{RG} H. E. Roman  and M. Giona,   {\it J. Phys. A: Math. Gen.} {\bf 25},
2107 (1992)
\bibitem{Wys} W. Wyss, J. Math. Phys. {\bf 27}, 2782 (1986).
\bibitem{SW} W. R. Schneider and W. Wyss, {\it J. Math. Phys.} {\bf 30}, 134 (1989).
\bibitem{Jum} G. Jumarie, {\it J. Math. Phys.} {\bf 33}, 3536 (1992).
\bibitem{Fog} H. C. Fogedby, {\it Phys. Rev. Lett.}{\bf 73}, 2517 (1994).
\bibitem{Zas} G. M. Zaslavsky, {\it Physica D} {\bf 76}, 110 (1994).
\bibitem{KG1} K. M. Kolwankar and A. D. Gangal, {\it Chaos} {\bf 6} 505 (1996).
(chao-dyn/9609016)
\bibitem{OS} K. B. Oldham  and J. Spanier,   {\it The Fractional Calculus}
 (Academic Press, New York, 1974).
\bibitem{RM} K. S. Miller  and B. Ross,  {\it An Introduction to the Fractional}
{\it Calculus and Fractional Differential Equations} 
(John Wiley, New York, 1993).
\bibitem{KG2} K. M. Kolwankar and A. D. Gangal, {\it Pramana - J. Phys.} 
{\bf 48}, 49 (1997).(chao-dyn/9711010)
\bibitem{Ris} H. Risken, {\it The Fokker-Planck Equation} (Springer-Verlag,
Berlin, 1984).
\bibitem{LM} E. W. Montroll and M. F. Schlesinger,
{\it On the wonderful world of random walks} in 
{\it Nonequlibrium Phenomenon II: From Stochastic to Hydrodynamics}
edited by J. L. Lebowitz and E. W. Montroll
(North-Holland, Amsterdam, 1984)
\bibitem{Fel2} W. Feller, {\it An Introduction to Probability Theory and its
Applications} (Wiley, New York, 1968) Vol 2.
\bibitem{KG3} K. M. Kolwankar  and A. D. Gangal,  (in preparation)
\bibitem{KG4} K. M. Kolwankar and A. D. Gangal, 
 in the proceedings of the
Conference `Fractals in Engineering', Archanon, 1997.(physics/9801010)
\end{thebibliography}
\end{document}